\documentclass[letter,preprint]{jpsj2}
\usepackage{graphicx}

\def\runauthor{Toshihiro \textsc{Kubo} and Yasuhiro \textsc{Tokura}}

\title{%
Many-Body Effects on Tunneling of Electrons in Magnetic-Field-Induced Quasi One-Dimensional Electron Systems in Semiconductor Nanowhiskers
}

\author{%
Toshihiro \textsc{Kubo}$^{1,}$\thanks{E-mail: j1202701@ed.kagu.tus.ac.jp} and Yasuhiro \textsc{Tokura}$^{2,}$\thanks{E-mail: tokura@will.brl.ntt.co.jp}
}

\inst{%
$^1$Department of Physics, Faculty of Science, Tokyo University of Science,\\
1-3 Kagurazaka, Shinjuku-ku, Tokyo 162-8601\\
$^2$NTT Basic Research Laboratories, NTT Corporation\\
3-1, Morinosato Wakamiya, Atsugi, Kanagawa, Japan
}

\recdate{\today}

\abst{%
Effects of the electron-electron interaction on tunneling in a semiconductor nanowhisker are studied in a magnetic quantum limit. We consider the system with which bulk and edge states coexist. In bulk states, the temperature dependence of the transmission probability is qualitatively similar to that of a one-dimensional electron system. We investigate contributions of edge states on transmission probability in bulk states. Those contributions can be neglected within our approximation which takes into account only most divergent terms at low temperatures.
}

\kword{%
magnetic quantum limit, Friedel oscillation, Hartree-Fock approximation, semiconductor nanowhisker, edge state
}

\begin{document}
\maketitle

\section{Introduction}
An isotropic bulk conductor placed in a very strong magnetic field, with only the lowest Landau subband occupied (\textit{magnetic quantum limit}, MQL) provides an interesting example of a quasi one-dimensional (1D) electron system. We thus expect its transport properties to be similar to those of 1D electron systems. Many-body effects on the electron transport in the magnetic-field-induced quasi 1D electron systems have recently been investigated\cite{maslov,glazman,kubo}. According to them, Friedel oscillations of the electron density induced by the barrier give an essential effect on the electron transport in magnetic-field-induced quasi 1D electron systems like the case of 1D electron systems. In such systems, measurement of the electron transport is much easier than in 1D electron systems, since it can be performed with use of bulk specimen.

We investigate effects of the electron-electron interaction on the transmission probability of electrons through a tunnel junction in a MQL. Starting with the Hartree-Fock theory, the Coulomb interaction, which gives rise to the divergence of Fock correction, should be replaced by the dynamically screened Coulomb interaction whereas we should use the bare Coulomb interaction for Hartree correction\cite{kubo,kawabata}. Nevertheless, the results obtained by the perturbation theory diverge logarithmically at low temperatures. So we take into account higher order contributions using the poor man's scaling approach\cite{poorman}. The temperature dependence of the transmission probability is qualitatively similar to that of a 1D Tomonaga-Luttinger liquid (TLL)\cite{matveev}, except that the parameter of the electron-electron interaction is magnetic field dependent, and may be either positive or negative. We show the magnetic field dependences of the parameter in some cases. The electron-electron interaction may either suppress or enhance the transmission, in contrast to TLL with the repulsive interaction.

Those predictions are experimentally verifiable by low carrier density materials, e.g. doped semiconductors. In order to observe clear interaction effects, in a MQL, the mean free path of the electrons has to be much longer than the Fermi wave length. However, for bulk doped semiconductors, it is difficult to satisfy that condition. Therefore, in this letter, we consider semiconductor nanowhiskers as more realistic systems because of the extremely high carrier mobility expected in modulation doped structures. Recently, high quality semiconductor nanowhiskers with sharp heterojunctions have been realized\cite{whisker1,whisker2}. Here we study interaction effects on the electron transport in a MQL in nanowhiskers whose radii are much longer than the Larmor radius. In such a system, there coexist bulk and edge states\cite{halperin}. We investigate contributions of edge states on the transmission probability in bulk states and show that those can be neglected within our approximation. Finally, we will discuss the temperature dependences of the conductance in the whole system.

\section{Model}
We consider the semiconductor nanowhisker whose radius is much longer than the Larmor radius $\lambda_B=\sqrt{\hbar/eB}$. We ignore the spin degree of freedom, for we assume that the spins of all electrons are completely polarized in the magnetic field. We choose the z-axis of the coordinate system along the magnetic field induced parallel to the growth direction of the nanowhiskers, and use the symmetric gauge $\mib{A}=(-By/2,Bx/2,0)$ for the vector potential $\mib{A}$.
\begin{figure}[htbp]
  \begin{center}
    \includegraphics[scale=0.3]{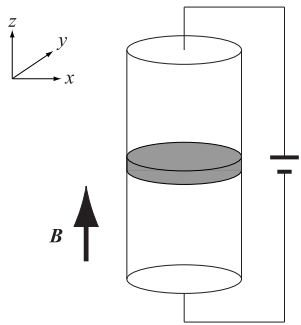}
  \end{center}
  \caption{Tunnel junction in a semiconductor nanowhisker whose radius is much longer than the Larmor radius. Magnetic field $\mib{B}=(0,0,B)$ is perpendicular to the tunnel barrier.}
    \label{nanowhisker}
\end{figure}
\begin{figure}[htbp]
  \begin{center}
    \includegraphics[scale=0.3]{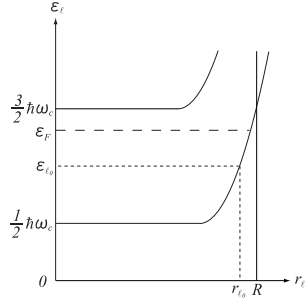}
  \end{center}
  \caption{Energy spectra in a plane perpendicular to a magnetic field. The radius $r_{\ell}$ is the center of the wave function. $\epsilon_F$ is the Fermi energy.}
    \label{spectrum}
\end{figure}

We model a tunnel junction in a nanowhisker (see Fig. \ref{nanowhisker}) by the following Hamiltonian:
\begin{equation}
H_0=\frac{\left(\mib{p}+e\mib{A}(\mib{x}) \right)^2}{2m}+W(x,y)+U(z),\label{hamiltonian}
\end{equation}
where $W(x,y)$ is the confining potential of a nanowhisker, and we approximate it by the infinite square well
\begin{align}
W(x,y)=\left\{
  \begin{array}{cc}
    0,   &  \sqrt{x^2+y^2}<R  \\
    \infty,   &  \sqrt{x^2+y^2}>R  \\
  \end{array}
\right.
,\label{potential}
\end{align}
where $R$ is the radius of the nanowhisker, $U(z)$ is the barrier potential, and we assume that the potential barrier is localized around $z=0$, i.e., $U(z)=0$ for $|z|>a$.

Moreover, in our systems, there coexist bulk and edge states as shown in Fig. \ref{spectrum}. Energy spectra in bulk states do not depend on the center of the wave function $r_{\ell}$ although those in edge states depend on $r_{\ell}$\cite{halperin}. Here, we do not consider the edge reconstruction predicted in two-dimensional systems\cite{edge1,edge2}.

First, we will consider effects of the electron-electron interaction on the transmission probability in bulk states. Next, we will investigate the contributions of edge states on them.

\section{Hartree-Fock Correction to Transmission Probability in Bulk States}
To begin with, we review the procedure of calculating interaction effects on the transmission probability in bulk states, where we assume that the electron-electron interaction is weak.

We consider the case where a MQL is realized. It is realized if the magnetic field is strong enough so that
\begin{equation}
B>\frac{\hbar}{e}(2\pi^4{n_e}^2)^{1/3},\label{mql}
\end{equation}
where $n_e$ is the electron density. Then it is well-known that the wave functions and energy eigenvalues in our systems are
\begin{subequations}
\begin{align}
&\varphi_{\ell,k_z}^{(0)}(\mib{x})=\phi_{\ell}^{(0)}(r,\theta)u_{k_z}^{(0)}(z),\label{wf}\\
&\epsilon_{k_z}^{(0)}=\frac{1}{2}\hbar\omega_c+\frac{\hbar^2{k_z}^2}{2m},\label{ee}
\end{align}
\end{subequations}
where
\begin{subequations}
\begin{equation}
\phi_{\ell}^{(0)}(r,\theta)=\frac{1}{\sqrt{2\pi 2^{|\ell|}|\ell|!}\lambda_B}\left(\frac{r}{\lambda_B} \right)^{|\ell|}e^{-\frac{r^2}{4{\lambda_B}^2}}e^{i\ell\theta}\quad ,\quad \ell\le 0,
\end{equation}
\begin{align}
u_{k_z}^{(0)}(z)=\left\{
  \begin{array}{cc}
   e^{ik_zz}+r_0e^{-ik_zz},& (z<-a),\\
   t_0e^{ik_zz},& (z>a),
  \end{array}
\right.\ (k_z>0),
\label{uk1}
\end{align}
\begin{align}
u_{k_z}^{(0)}(z)=\left\{
  \begin{array}{cc}
    t_0e^{ik_zz},& (z<-a),   \\
    e^{ik_zz}+r_0e^{-ik_zz}, & (z>a),
  \end{array}
\right.\ (k_z<0),
\label{uk2}
\end{align}
\end{subequations}
where $\omega_c=eB/m$ is the cyclotron frequency and $t_0$ and $r_0$ are the transmission and reflection amplitudes through the barrier, respectively.

\begin{figure}[htbp]
  \begin{center}
    \includegraphics[scale=0.5]{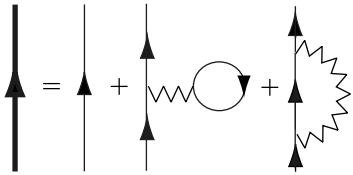}
  \end{center}
  \caption{Feynman diagram for Matsubara Green's function $\mathcal{G}$ to the lowest order in the interaction within the Hartree-Fock approximation. The thin solid lines indicate the Green's function for bulk states without the interaction and the wavy lines indicate the interaction.}
    \label{hf1}
\end{figure}

First, we investigate the correction to the transmission probability due to the electron-electron interaction to the lowest order within the Hartree-Fock approximation. We calculate the correction to the wave functions using Green's function methods\cite{agd}, where we use Matsubara Green's function $\mathcal{G}(\mib{x},\mib{y},\omega_n)$ whose Feynman diagrams are depicted in Fig. \ref{hf1}. As for the detailed calculations, the reader is referred to Ref. \citen{kubo}. The first-order correction to the wave function due to the interaction is given by
\begin{align}
\varphi_{\ell,k_z}^{(1)}(\mib{x})&=\int d{\bf x}_1\int d{\bf x}_1'G_{k_z}({\bf x};{\bf x}_1)
\left[\delta({\bf x}_1-{\bf x}_1')V_H({\bf x}_1')\right.\nonumber\\
&\left. -V_F({\bf x}_1,{\bf x}_1') \right]\varphi_{\ell,k_z}^{(0)}({\bf x}_1'),\label{correction}
\end{align}
where $G_{k_z}({\bf x};{\bf x}_1)$ is the single-electron retarded Green's function in a MQL and $V_H$ and $V_F$ are the Hartree and Fock potentials, respectively.

It is to be noted that if the interaction is the bare Coulomb interaction, Fock term in Eq. (\ref{correction}) is divergent because of its long-range property. According to Refs. \citen{kubo} and \citen{kawabata}, the interaction should be replaced by the screened interaction within random phase approximation. The replacement of the interaction for Hartree term is not appropriate, however, as is shown in Ref. \citen{kubo}. 
Thus we should keep the bare Coulomb interaction for Hartree term.The 1D-like Friedel oscillation of the electron density in a MQL leads to a logarithmic singularity in $\varphi_{{\bf k}}^{(1)}({\bf x})$. For electrons near the Fermi wave number $k_F=\sqrt{2m(\epsilon_F-\hbar \omega_c/2)}/\hbar$, we have the 1st-order correction to the transmission probability
\begin{equation}
\mathcal{T}^{(1)}=-2\alpha(B)\mathcal{T}_0(1-\mathcal{T}_0)\ln\left(\frac{E_0}{E} \right),\label{transmission}
\end{equation}
where $\mathcal{T}_0$ is the bare transmission probability, $E_0$ is the effective bandwidth, $E$ is the electron energy measured from Fermi energy in the linearized dispersion, and $\alpha(B)$ is the parameter of the electron-electron interaction given by
\begin{equation}
\alpha(B)=2(\kappa \lambda_B)^2\int^{\infty}_{0}dq\frac{qe^{-q^2{\lambda_B}^2/2}}
{q^2+\kappa^2e^{-q^2{\lambda_B}^2/2}}-\frac{\kappa^2}{(2k_F)^2},\label{parameter}
\end{equation}
where $\kappa$ is defined by
\begin{equation}
\kappa\equiv\sqrt{\frac{e^2}{4\pi^2\epsilon\hbar v_F{\lambda_B}^2}}.
\end{equation}
We have to take into account the higher order contributions in the interaction since Eq. (\ref{transmission}) is no longer valid at low temperatures.

In order to include the higher order contributions, we use the poor man's scaling approach\cite{poorman}. 
Then the following renormalization group equation is obtained
\begin{equation}
\frac{d\mathcal{T}}{d\ln(E_0/E)}=-2\alpha(B)\mathcal{T}(E)(1-\mathcal{T}(E)).\label{rg}
\end{equation}
From Eq. (\ref{rg}), the transmission probability becomes
\begin{equation}
\mathcal{T}(T)=\frac{\mathcal{T}_0(k_BT/E_0)^{2\alpha(B)}}
{\mathcal{R}_0+\mathcal{T}_0(k_BT/E_0)^{2\alpha(B)}}\label{temperature}
\end{equation}
where $\mathcal{R}_0=1-\mathcal{T}_0$. This result is of the same form as that of 1D electron systems except for the parameter of the electron-electron interaction. This is the transmission probability per mode. Therefore, from the Landauer formula, the total tunneling conductance in bulk states is
\begin{equation}
G_{bulk}(T)=\frac{e^2}{h}\frac{R^2}{2{\lambda_B}^2}\frac{\mathcal{T}_0(k_BT/E_0)^{2\alpha(B)}}
{\mathcal{R}_0+\mathcal{T}_0(k_BT/E_0)^{2\alpha(B)}},\label{landauer}
\end{equation}
where the factor $R^2/2{\lambda_B}^2$ is the number of bulk modes.
\begin{figure}[htbp]
  \begin{center}
    \includegraphics[scale=0.5]{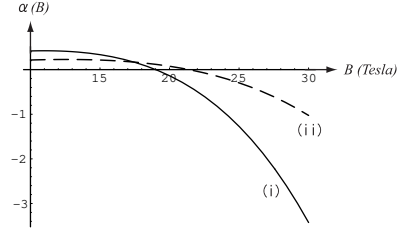}
  \end{center}
  \caption{The examples of the parameter $\alpha(B)$ vs. magnetic field $B$ for $n_e=10^{23} m^{-3}$, where the MQL is realized for $B>8.2 \ \mbox{Tesla}$. (i) $\epsilon=13.1\epsilon_0, m=0.067m_e$ (typical values of GaAs) and (ii) $\epsilon=14.55\epsilon_0, m=0.026m_e$ (typical values of InAs), where $\epsilon_0$ is the permittivity of vacuum and $m_e$ is the rest mass of electron.}
    \label{parameter}
\end{figure}

From Eq. (\ref{temperature}), we find that the transmission probability vanishes as $T\to 0$ if $\alpha(B)>0$, as in the case of 1D systems with the repulsive interaction. However $\alpha(B)$ can be negative for large magnetic fields (see Fig. \ref{parameter}). In this case, according to Eq. (\ref{temperature}), the transmission probability should become unity at zero-temperature.
In Fig. \ref{parameter}, magnetic field dependences of the parameters are shown for the standard doped semiconductor, GaAs, and InAs nanowhiskers realized in Refs. \citen{whisker1} and \citen{whisker2}. 

\section{Contributions of Edge States on Transmission Probability in Bulk States}
\begin{figure}[htbp]
  \begin{center}
    \includegraphics[scale=0.5]{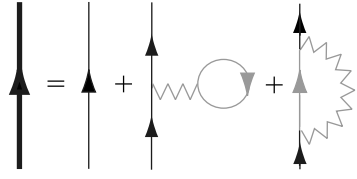}
  \end{center}
  \caption{Feynman diagram for Matsubara Green's function as the contribution of edge states to bulk states: The thin gray lines indicate the Green's function for edge states without the interaction.}
    \label{hf2}
\end{figure}

Here the contributions of edge states on the transmission probability in bulk states are studied. For our choice of the gauge, energy spectra in edge states are dependent on the center of the wave function $r_{\ell}$, as shown in Fig. \ref{spectrum}. Therefore, the Fermi wave number $k_F'$ in the $z$-direction in edge states is smaller than the Fermi wave number $k_F$ in the $z$-direction in bulk states. We study the contributions of the edge state for the angular momentum $\ell=\ell_0$ (see Fig. \ref{spectrum}). Then we have
\begin{equation}
k_F'=\frac{\sqrt{2m\left(\epsilon_F-\epsilon_{\ell_0} \right)}}{\hbar},\label{fermi}
\end{equation}
where $\epsilon_F$ is the Fermi energy and $\epsilon_{\ell_0}$ is the energy of the edge state with $\ell=\ell_0$. Here we calculate Matsubara Green's function for the Feynman diagram shown in Fig. \ref{hf2} as the contribution of edge states to bulk states. Then the first-order correction to the transmission probability in bulk states interacting with the edge state is obtained by
\begin{equation}
\mathcal{T}'^{(1)}=-2\alpha'(B)\mathcal{T}_0(1-\mathcal{T}_0)\ln\left(\frac{1}{|k_z-k_F'|d} \right),\label{edge}
\end{equation}
where $\alpha'(B)\neq\alpha(B)$, $\alpha'(B)$ can be expressed using wave functions of the edge state, and $d\equiv \hbar v_F/E_0$ is the cut-off length. This equation is derived from the same formulation as in Sec. 3. Here $k_z$ is the wave number of electrons in bulk states and only electrons of $k_z\simeq k_F$ contribute electric conduction at low temperatures. Since we take into account only most divergent terms at low temperatures, this correction can be ignored because of $k_F'\neq k_F$. Therefore we find that the logarithmic singularity does not appear in this case. As a result, even if we take the scattering from edge states into consideration, Eq. (\ref{landauer}) is still valid at low temperatures. Since the wave vector $2k_F'$ of the Friedel oscillation of the electron density in the edge state does not match the momentum transfer $2k_F$ in bulk modes, this result arises.
Provided we may take into account the higher order corrections, the contributions from the edge states can be ignored within the approximation which takes the most divergent term in the $n$th order of the form $[\alpha\ln(1/|k_z-k_F|d)]^n$.

\section{Discussion}
\begin{figure}[htbp]
  \begin{center}
    \includegraphics[scale=0.4]{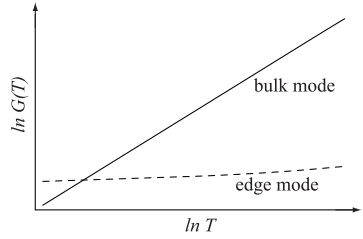}
  \end{center}
  \caption{Conductance vs. temperature. Each curve of the temperature dependences of the conductances in bulk and edge modes is the case when the parameters $\alpha(B)$ and $\alpha_{e,\ell}(B)$ are positive.}
    \label{conductance}
\end{figure}

In previous sections, we considered only tunneling in bulk states. However, tunneling of edge states may not be disregarded in the actual experiments. Although contributions of both bulk and edge states are observed experimentally, we anticipate that singular behaviors of the transmission probability in bulk states can be observed for the following reasons. (i) In the case where the scattering between bulk and edge states can be ignored, the total conductance is given by $G_{tot}=G_{bulk}+G_{edge}$. We can easily estimate $G_{edge}/G_{bulk}\simeq \lambda_B/R$ even if the conductance per mode is the same order of magnitude. Therefore, if $R\gg \lambda_B$, contributions due to tunneling of edge states are very small. (ii) The bare transmission probability in the edge state is smaller than that in the bulk state because of $k_F'<k_F$. For a delta-function like potential, $U(z)=\hbar^2 Q\delta(z)/2m$, the transmission probability of an electron with wave number $k_z$ is given by $1/(1+(Q/2k_z)^2)$. (iii) Since the Fermi wave number in the edge states differs for every mode, only in the case when the electron in the edge state is scattered on the density modulation due to its own mode, logarithmic singularity arises. In contrast, the bulk states suffer scattering by the density modulation in all modes showing $2k_F$ oscillations. Consequently, we expect that the absolute values of the parameters of the interaction are small, i.e., $|\alpha_{e,\ell}(B)|<|\alpha(B)|$, where $\alpha_{e,\ell}(B)$ is the parameter of the interaction related to the tunneling in a single edge state $\ell$. The temperature dependences of conductances in bulk and edge states are schematically shown in Fig. \ref{conductance}. The slopes of curves in Fig. \ref{conductance} indicate parameters $\alpha(B)$ and $\alpha_{e,\ell}(B)$. Therefore, at high temperatures, the temperature dependence of the conductance in bulk states is dominant, and the conductance in edge states hardly depends on temperature. At low temperatures, if the transmission probability in bulk states approaches $0$ or $1$, the temperature dependence of the conductance will be determined only by the  contribution of edge states, and the temperature dependence is dominated by the edge state with smallest $|\alpha_{e,\ell}(B)|$. 
The edge states (dotted line) do not obey a power-law behavior since modes with different $\alpha_{e,\ell}(B)$'s contribute to the conductance.

In a MQL, however, it is well known that the electron-electron interaction leads to a charge density wave (CDW) instability of the ground state\cite{fukuyama}. Therefore, our discussions are valid at higher temperatures than the CDW transition temperature. 

\section{Summaries}
We have shown that the electron transport in magnetic-field-induced quasi 1D electron systems is qualitatively similar to that of 1D electron systems. However, since parameters of the interaction are magnetic field dependent, we obtained results quantitatively different from those of 1D electron systems. Moreover, we have extended our theory to semiconductor nanowhiskers which are more realistic systems, and have shown that contributions of edge states on the transmission probability in bulk states can be ignored within our approximation.  

\section*{Acknowledgments}
One of the authors (Y. T.) is partly supported by SORST-JST.

\end{document}